\documentclass[journal]{IEEEtran}
\usepackage{cite}
\usepackage{amssymb}
\usepackage{graphicx}
\usepackage{multirow}
\usepackage{subfigure}
\usepackage{rotating}
\usepackage{multirow}
\usepackage{booktabs}
\usepackage{makecell}

\usepackage{color}
\newcommand{\tabincell}[2]{\begin{tabular}{@{}#1@{}}#2\end{tabular}}
\begin{document}
%\title{An End-to-End neural network frame for sparse-views CT reconstruction}
\title{Learning image from projection: a full-automatic reconstruction (FAR) net for sparse-views computed tomography
	\thanks{This work was partly supported by the National Natural Science Foundation of China (grants: 61827809, 61671311, 61501310, 61371195), Beijing Commission of Education grant KM201710028002.}
	\thanks{G. Ma, Y. Zhu, X. Zhao are with school of Mathematical Sciences, Capital Normal University, Beijing, 100048, China, and with Beijing Advanced Innovation Center for Imaging Technology, Capital Normal University, Beijing, 100048, China. e-mail: (MaGenwei@126.com, ynzhu@cnu.edu.cn, zhaoxing\_1999@126.com).}
	\thanks{Y. Zhu serves as the correspondence author (e-mail: ynzhu@cnu.edu.cn).}}
\author{Genwei Ma, Yining Zhu$^\star$, \IEEEmembership{Member, ~IEEE}, Xing Zhao}
\maketitle
\IEEEpeerreviewmaketitle
\begin{abstract}
The sparse-views x-ray computed tomography (CT) is essential for medical diagnosis and industrial nondestructive testing. However, in particular, the reconstructed image usually suffers from complex artifacts and noise, when the sampling is insufficient. In order to deal with such issue, a full-automatic reconstruction (FAR) net is proposed for sparse-views CT reconstruction via deep learning technique. Different with the usual network in deep learning reconstruction, the proposed neural network is an end-to-end network by which the image is predicted directly from projection data. The main challenge for such a FAR-net is the space complexity of the CT reconstruction in full-connected (FC) network. For a CT image with the size $N \times N$ , a typical requirement of memory space for the image reconstruction is $O (N^{4})$, for which is unacceptable by conventional calculation device, e.g. GPU workstation. In this paper, we utilize a series of smaller FC layers to replace the huge  layer based on the sparse nonnegative matrix factorization (SNMF) theory. By applying such an approach, the FAR-net is able to reconstruct sparse-views CT images with the size $512\times 512$ on only single workstation. Furthermore, an artifact suppression neural network (AS-NN) structure is composed in the FAR-net for suppressing the artifacts and noise caused by under-sampling data. The results of numerical experiments show that the projection matrix and the FAR-net is able to reconstruct the CT image from sparse-views projection data with a superior quality than conventional methods such as FBP and optimization based approach. Meanwhile, the factorization for the inverse projection matrix is validated in numerical.
\end{abstract}

\begin{IEEEkeywords}
deep learning, sparseness non-negative matrix factorization, Full-automatic reconstruction, sparse-views CT
\end{IEEEkeywords}
\IEEEpeerreviewmaketitle
\section{Introduction}
\IEEEPARstart{X}-ray computed tomography (CT) has been widely used in medical diagnostic and industrial nondestructive testing due to its great ability in visualizing interior structure. Considering the damage of radiation to patients, it is of great significance and essential to reduce the dose in the practical application. One of the effective methods to reduce the dose is to decrease the scanning angles which be named sparse-views CT. However, the sparse-views scanning will lead in a variety of issues in reconstructed images such as artifacts and noise. Many studies have demonstrated that the traditional methods \cite{Zhang2018Limited}, such as filter back-projection (FBP), algebraic reconstruction algorithm (ART) \cite{Gordon1970Algebraic}, simultaneous algebraic reconstruction technique (SART) \cite{Andersen1984Simultaneous}, expectation maximization (EM) \cite{Dempster1977Maximum}, are failed to deal with these issues via theories or experiments. In order to improve the quality of reconstructed images from sparse-views scanning, various methods have been proposed according to compressive sensing (CS) theory\cite{Donoho2006Compressed}. Most of them are based on optimization model which imports some prior knowledge as constraints terms. For example, some studies utilize $L_1$ regularization term of the gradient of images for dealing with different issues, such as sparse-views, low-dose,  limited-angle as well as interior tomography. Such type of regularization is also named as Total variation (TV) \cite{Sidky2008Image, Zhang2014Few, Zhang2016Statistical, Zhang2013Few} which is based on the assumption that the gradient of CT images is sparse. Inspired by the sparsity of image, many approaches are proposed e.g. dictionary learning \cite{Xu2012Low}, non-local means (NLM) \cite{Chen2009Bayesian, Jianhua2012Iterative, Zhang2016Spectral} and different wavelet transform methods \cite{Verma2015Denoising}\cite{Kallel2018CT}. These approaches prove satisfied results in varying degrees for removing artifacts and suppressing noise. However, it still remain two main barriers for applying the approach to practical application: the great computation cost caused by the iterative manner of the solving algorithm \cite{Wang2017Local} and the difficulty of choosing the regularization parameters. Hence, it would be desperately desire for an approach with low complicated of computation and convenient parameter choosing in practical application.

Recently, deep learning technology has been applied in many fields with remarkable results which is benefited from dramatic improvements in CUDA acceleration \cite{CUDA_dl}. In the field of CT image processing, there are various convolution neural network (CNN) and recurrent neural network (RNN)  been proposed for image restoration \cite{Zhang2017Learning}\cite{Yu2018Crafting}, denoising, artifacts correction \cite{Zhang2017Convolutional}\cite{Gjesteby2017Deep}, low-dose CT \cite{Chen2017Low, Kang2017A, Chen2017Lowdose, DBLP:journals/corr/abs-1708-00961}, sparse-views CT \cite{Han2017Framing} and spectral CT \cite{Zhang2018Image} in image domain. Furthermore, researchers began to incorporate iterative method into CNN \cite{Chen2018LEARN}\cite{Gong2017Iterative}. Although the results reported so far are remarkable in terms of image quality, these methods and the corresponding CNN frame can only be utilized in image domain, which is post-processing without achieving the stage of image reconstruction. In other words, these methods are semi-traditional and semi-deep learning. However, a deep network mimics an organism better than a linear operator, and is much more intelligent than a linear system solver \cite{Perspective} as well. Hence, in practice, it is of great importance  to build a end-to-end network which try to translate an image in a modality that is difficult to understand to a corresponding image which can be recognized by human \cite{Liu2017Unsupervised}\cite{Yi2017DualGAN}. W$\ddot{u}$rfl \emph{et al}. have already demonstrated that image reconstruction can be expressed in term of neural network and shown that FBP can be mapped identically onto a deep neural network architecture \cite{W2016Deep}. In this method,  the parameters of fully connected (FC) layer of neural network are fixed and pre-calculated by discrete formulation of FBP algorithm and the parameters of convolution layer are initialized by Ram-Lak filter. Argyrou \emph{et al}. proposed an approach of artificial neural network reconstruction, and Zhu \emph{et al}. show that sensor domain to image domain via automated transform by manifold approximation (AUTOMAP) \cite{Zhu2018Image}. Such neural network require an impractical amount of  memory space which hamper the approach applied to practical. More specifically, the proposed neural network needs to train huge amount of parameters. The space complexity the of the network parameters is $O (n^4)$, where $n$ is the width or height of reconstruction image. Because of memory limitation, the scheme can only reconstruct low resolution images. For example, if we want to reconstruct an image with a resolution of $512\times 512$ using AUTOMAP method, the number of parameters (single float) is $512^4$, requiring 1 TB memory for which will be a heavy cost of computational hardware cost such as GPU. Li \emph{et al}. proposed iCT-Net for CT images from sinogram data\cite{iCTNet}, which achieve back projection by introducing a fixed and separated rotation layer which is calculated based on the CT system's configuration. By this strategy, the size of mapped network could decrease dramatically. 

In this paper, in order to take advantage and overcome shortcomings of end-to-end neural network for image reconstruction, we propose an approach to reduce the size of the FC layer based on the idea of sparseness non-negative matrix factorization (SNMF) \cite{hoyer2004non}. Non-negative matrix factorization (NMF) is usually utilized for parts-based representations in machine learning \cite{lee1999learning} in earlier study.For a given non-negative data matrix $V$, NMF finds an approximate factorization $W \approx V\cdot H$ into non-negative factors $V$ and $H$\cite{NIPS2000_1861}. Especially, if we make sparseness constraints at the matrix, the better quality will be achieved for NMF. By applying the SNMF idea, the space complexity of the network has been reduced to $O (n^3)$ and the requirement of the memory space has been decreased to an acceptable size. For example, we are even able to reconstruct CT image with a resolution of $512\times 512$ for sparse-view scanning in a single  workstation with multi-GPUs. Moreover, we integrate an artifact suppression neural network (AS-NN)  to our network for furtherly suppressing the residual artifacts and noise caused by the sparse-views scanning. Hence, our network would learn features from data and reconstruct image without manual intervention and we name it as Full-Automatic Reconstruction (FAR) net. The numerical experiments show that FAR-net could be implemented on only one workstation and predict the CT image from projection data directly with a superior quality than traditional algorithm such as TV based approach.

The remain of the paper is organized as follows: the section \ref{Pk} introduce related knowledge of CT as well as neural network  including FC network and CNN. In the section \ref{Mt}, we describe the proposed FAR-net which contains  two strategies: reconstruction and  artifacts suppression. In the section \ref{NE}, numerical experiments are carried out to verify the CT problem can map into FC network in low dimensions, but also is useful for sparse-views CT problem by two neural network. Finally, we summarize the paper in section \ref{summary}.
\section{Preliminary knowledge}\label{Pk}
\subsection{CT imaging}
In ideal condition, the mathematical model of CT is usually described as a discrete linear system \cite{Zhao2015An}
\begin{equation}\label{AX_b}
  AX = b
\end{equation}
where $A\in\mathbb{R}^{M\times N}$ denotes the projection matrix, $\textbf{x}$ denotes the reconstructed image  and $b$ denotes the projection data. It is an inverse problem for solving  $\textbf{x}$ from $b$. For such a problem, it is difficult to obtain $\textbf{x}$ directly since the system matrix $A$ is too huge to find the inverse matrix. Here the system matrix is $A_{ (u\times v)\times (w\times h)}\triangleq A_{M\times N}$  where $u$, $v$ are the number of detector bins and scanning samples, and $w$, $h$ are the size of reconstruction image. It is noteworthy that $A$ would be a typical sparse non-negative matrix and the size of would achieve about $2^{18}\times 2^{15}$ for a typical medical image.
%Since 
%The elements of $A$ are the length of the interaction  between x-rays and image pixels. However, it is only exceed less than $2\sqrt{w \times h }$ points of discrete grid when penetrating a specimen for each x-ray. Hence, 

\subsection{Neural network}
In recent years, the deep  neural network (NN) has sprung up rapidly and is widely used in classification, object detection \cite{Girshick2015Fast, Liu2016SSD}, image segmentation \cite{DBLPJegouDVRB16, DBLPRonnebergerFB15} and so on. Fully connection neural network (FCLs) and convolutional neural network (CNN) are two typical deep neural network. In this subsection, we will introduce the principle of these two networks.
\subsubsection{Fully connection neural network}
\begin{figure}[htbp]
  \centering
  \includegraphics[width = 3in]{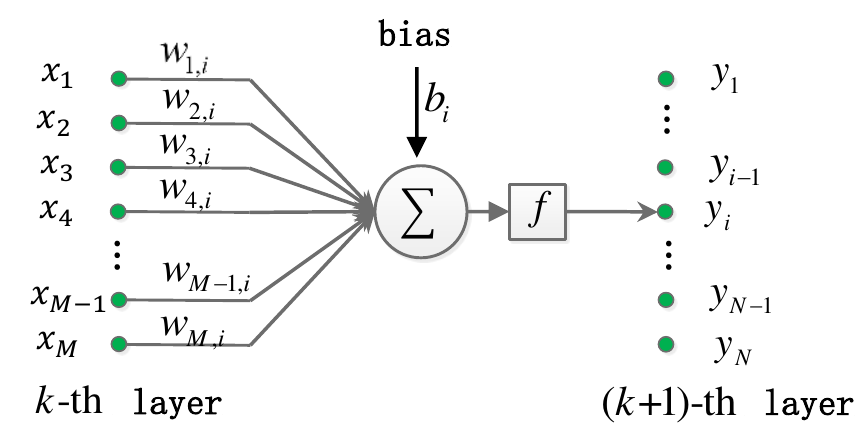}
  \caption{A fully-connected layer between $k$-th layer and $ (k+1)$-th layer.}\label{full_conn}
\end{figure}

As shown in Fig. \ref{full_conn}, the $k$-th layer ($X= (x_1, x_2, \cdots, x_M)^T$) has $M$ neurons and the $ (k+1)$-th layer ($Y= (y_1, y_2, \cdots, y_N)^T$) has $N$ neurons. So the weight matrix ($W$) of the two adjacent fully connected layers is a $M\times N$ matrix. A FCL can be described as Eq. (\ref{a_connected}):
\begin{equation}\label{a_connected}
  Y = f (W\cdot X+b)
\end{equation}
where $f$ is the activation function and $b$ is the bias. Without considering the bias and activation function, a FCL can be expressed as

\begin{equation}
Y = W\cdot X.
\end{equation}
which is very similar to Eq. (\ref{AX_b}).

\subsubsection{Convolutional neural network}
Various types of CNNs have been proposed such as ResNet \cite{DBLPHeZRS15}, U-net \cite{DBLPRonnebergerFB15}, DenseNet \cite{DBLHuangLW16a}. The simplest form of CNNs with $n$ layers is expressed as
\begin{equation}\label{CNN}
\begin{array}{rcl}
  Y =f_n (BN (W_n*f_{n-1} (\cdots (W_2*f_1 (BN (W_1*X\\+b_1)+b_2))\cdots)+b_n))
  \end{array}
\end{equation}
where $X$ is the input data, $W_i$ is the convolution kernel of the $i$-th layer, $b_i$ is the bias of the $i$-th convolution layer, $f_i$ is an activation function, $*$ is represented the convolution operation,  $BN$ denotes batch normalization and $Y$ is output data or prediction. The goal of CNN framework is to find optimal parameters $W_i, b_i (i=1, 2, \cdots, n)$ to minimize the following energy function, 
\begin{equation}\label{eFunc}
E = ||Y^{\star} - Y||_{L_{p}}
\end{equation}
where $Y^{\star}$ is the ground truth data and $L_{p}$ is norm to measure the difference between the predicted result and the real data.

%Our neural network consists of two functions: reconstruction and suppressing artifacts. 

\section{Method}\label{Mt}
As mentioned above, it is an enormous challenge to reconstruct CT image  from projection data via a deep neural network without any manual intervention due to the unacceptable requirement of memory space. 
In this section, inspired by the matrix factorization, we propose a Full-Automatic Reconstruction (FAR) net to predict CT image from spare-view projection data directly on a single workstation. The FAR-net is motivated by the following observations: 
\begin{itemize}
\item the sparse matrix can be decomposed into the product of a series of matrices approximately based on SNMF theory; 
\item sparse-views CT images usually suffer from heavy artifacts and noise, and CNNs has great potential to remove such artifacts and noisy;  
\item  a deep neural network is ideal to capture various types information from a large amount of training data \cite{Kang2017A}. 
\end{itemize}
Hence, we designed the FAR-net to provide two functions including image reconstruction image and artifacts suppression via FCLs and CNNs.
\subsection{Reconstruction Neural Network}
According NMF theory \cite{hoyer2004non}, a given non-negative matrix ($W$) can be factorized as follow:
\begin{equation}\label{SNMF}
  W_{M\times N} \approx V_{M\times c}H_{c\times N}
\end{equation}
where $c<min (M, N)$ and both $V_{M\times c}$ and $H_{c\times N}$ are non-negative. 
To improve the quality of approximation of Eq. \ref{SNMF}, there are different cost functions such as $L_{2}$ norm or Kullback-Leibler divergence:

\begin{equation}\label{L2}
E_{L_{2}}=||VH - W||^{2}_{2}, 
\end{equation}
\begin{equation}\label{K-L}
E_{KL}=\Sigma_{ij}[ (V H)_{ij}\log\frac{ (VH)_{i, j}}{W_{ij}}- (VH)_{ij} + W_{ij}].  
\end{equation}
Lee \emph{et al.} have found an algorithm to minimize  $E_{L{2}}$ and $E_{KL}$ and gave the proof of convergence \cite{lee1999learning}. Furthermore, Hoyer \emph{et al.} indicated explicitly that incorporating the sparseness as contrast for matrix could improve the result of factorization \cite{hoyer2004non}. If we decompose $V$ and $H$ continually, the  theory still works since both of them are SNMF. Hence, we could factorize the SNM within a few steps, e.g. 2 or 3 layers. Then, considering the huge projection matrix $A$ in Eq.\ref{AX_b}, it could be approximatively represented by a serious of smaller SNMs as follows, 
\begin{equation}\label{AWWW}
A \approx V_{1}\cdots V_{n}\cdot H_{1}\cdots H_{n}.
\end{equation}

\begin{figure}[htbp]
  \centering
  \includegraphics[width = 3.2in]{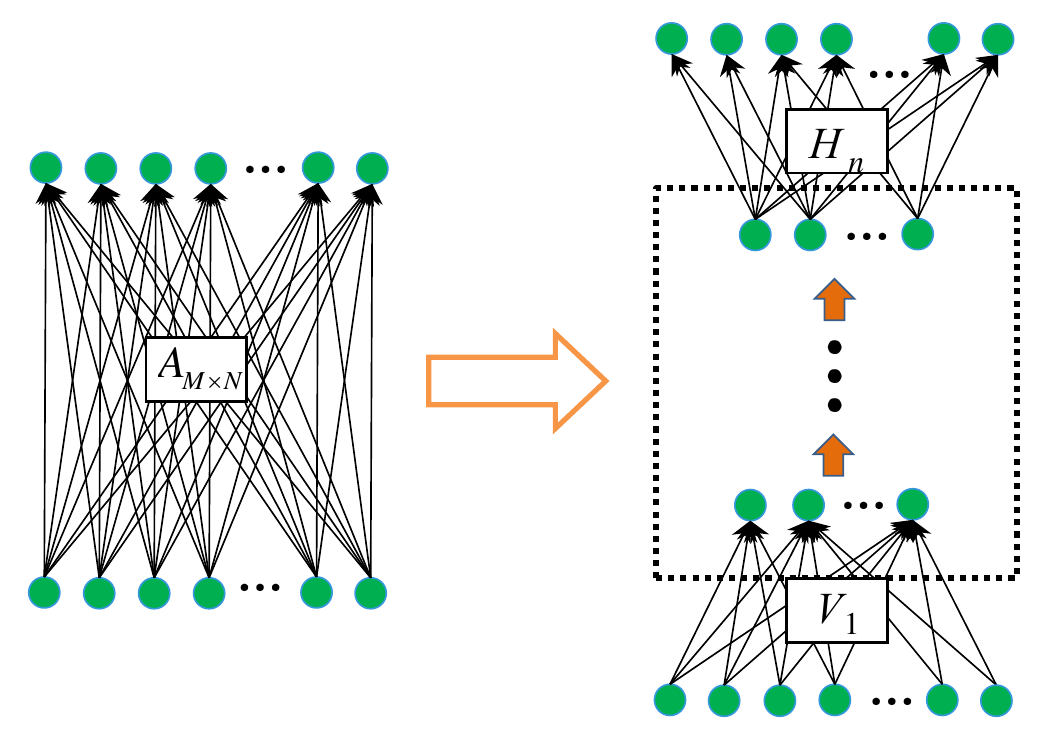}
  \caption{Mapping matrix factorization  into FC network. A inverse of SNM can be factor as some smaller matrix.}\label{SNMF_layer}
\end{figure}

\begin{figure*}[htbp]
  \centering
  \includegraphics[width = 7in, , height=3in]{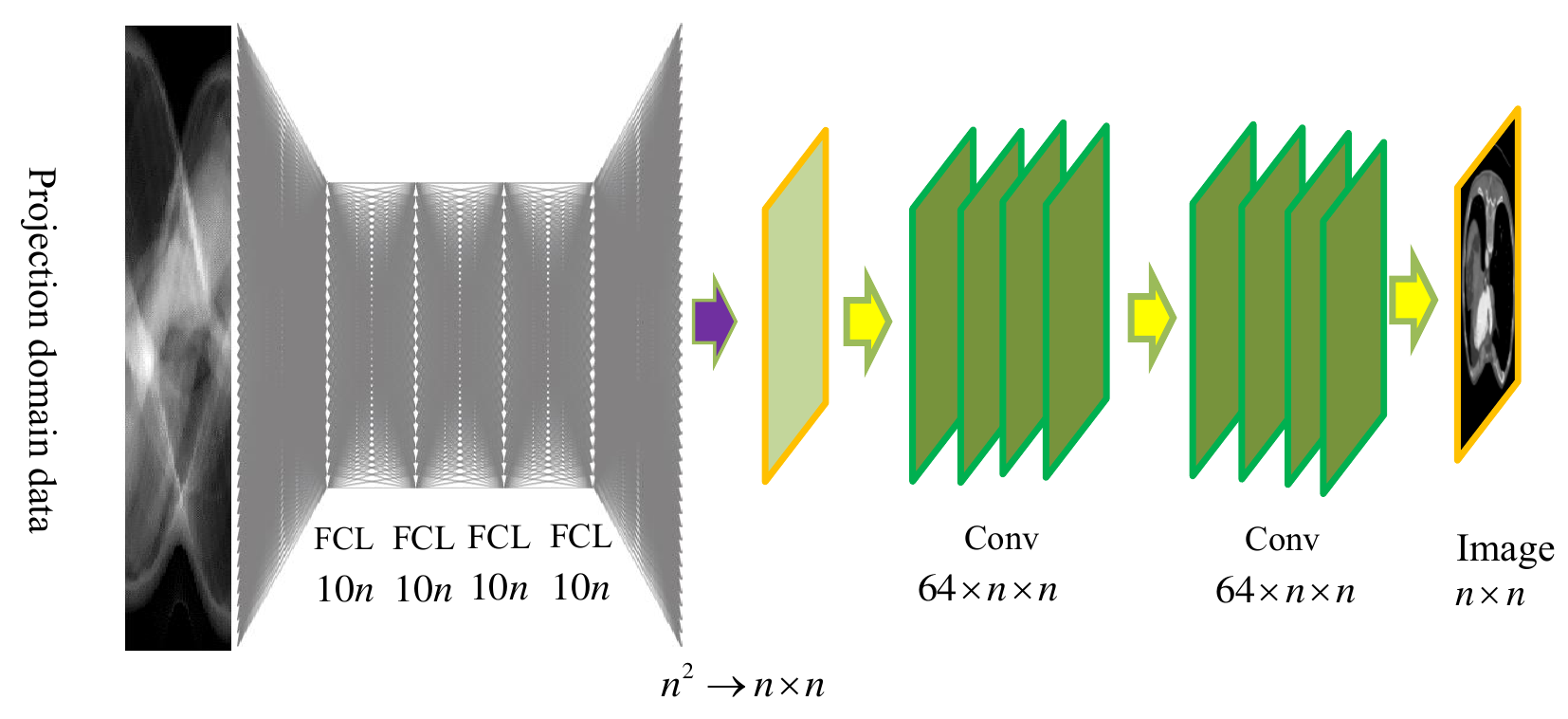}
  \caption{Recon-NN frame for image reconstruction strategy. There are two parts in the frame. The first part consist five FC layers and the second part include three convolutional layers. }\label{fcl_layer}
\end{figure*}

In particular, the CT reconstruction can be considered as two processes: filter and back projection (FBP). The process of back projection also can be regarded as a linear transformation, and the sparse property of this linear transformation matrix $R$ is consistent with the projection matrix A. The process of filter can be realized by convolution layer. Hence, we are able to learn the inverse of $R$ based on Eq. \ref{AWWW} via FCLs neural network which are composed by some smaller middle layers. As shown in Fig. \ref{SNMF_layer}, these middle layers can effectively reduce the amount of network parameters and the requirement of memory space. Based on such a structure, we proposed the reconstruction neural network shown in Fig. \ref{fcl_layer} (Recon-NN). Although Eq. \ref{AWWW} is not strictly proofed in theory, numerical experiments indicate that the weights matrices of trained network is approximately enough equal to the inverse matrix. Therefore, the network is able to predict the CT image from projection data directly.

\subsection{Artifacts suppression neural network}
\begin{figure*}[htbp]
\includegraphics[width=7in, height=3.5in]{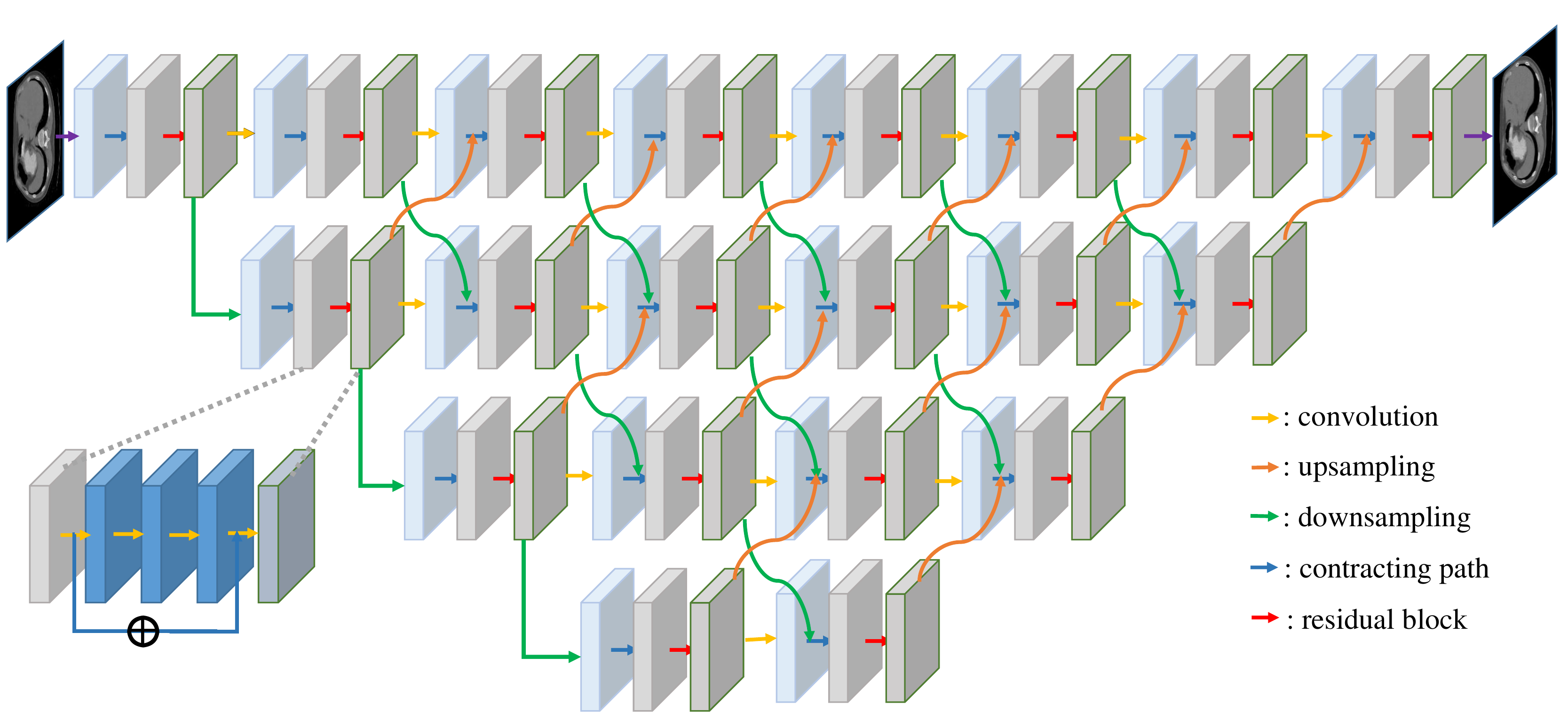}
\caption{AS-NN frame for artifact suppression and de-noising. The red arrow correspond to residual blocks and different scales of space are merged by convolution, up-sampling and down-sampling}\label{unet}
\end{figure*}

Generally speaking, sparse-views CT images usually suffer from complex artifacts as well as noise. In order to solve this issue, we propose an  artifacts suppression neural network (AS-NN) to improve the quality of image which is predicted from Recon-NN. As you can see in Fig. \ref{unet}, our network takes full advantage of residual block \cite{DBLPHeZRS15} and U-net architecture \cite{DBLPRonnebergerFB15}. More specifically, the residual block are shown in the lower left corner of the Fig. \ref{unet}. The bypass connections in the residual block is able to  recover image with higher quality and to avoid vanishing gradient problem in back-propagating. Similarly, The U-net architecture  also can preserves the details of high-frequency features. Since a typical CNN has pooling layers, the information may be lost after passing these layers. To avert this phenomenon, high-frequency features from the contracting path are combined with up-sampled output to recover the details\cite{Kang2017A}. In addition, our network has added its own features. Firstly, our network performs down-sampling and up-sampling on different scale spaces to realize information exchange in different scale spaces, while U-net has less communication between data in different scale spaces, and there is only one down-sampling and up-sampling in adjacent scales. Secondly, the information contained in different scale spaces is different. The space with larger resolution should contain more information, The space with larger resolution should contain more information, and the network should be set deeper. The number of convolution layers of our network for different scale spaces is different, while unet is the same different scale spaces.

\begin{figure*}[htbp]
\includegraphics[width=7.0in]{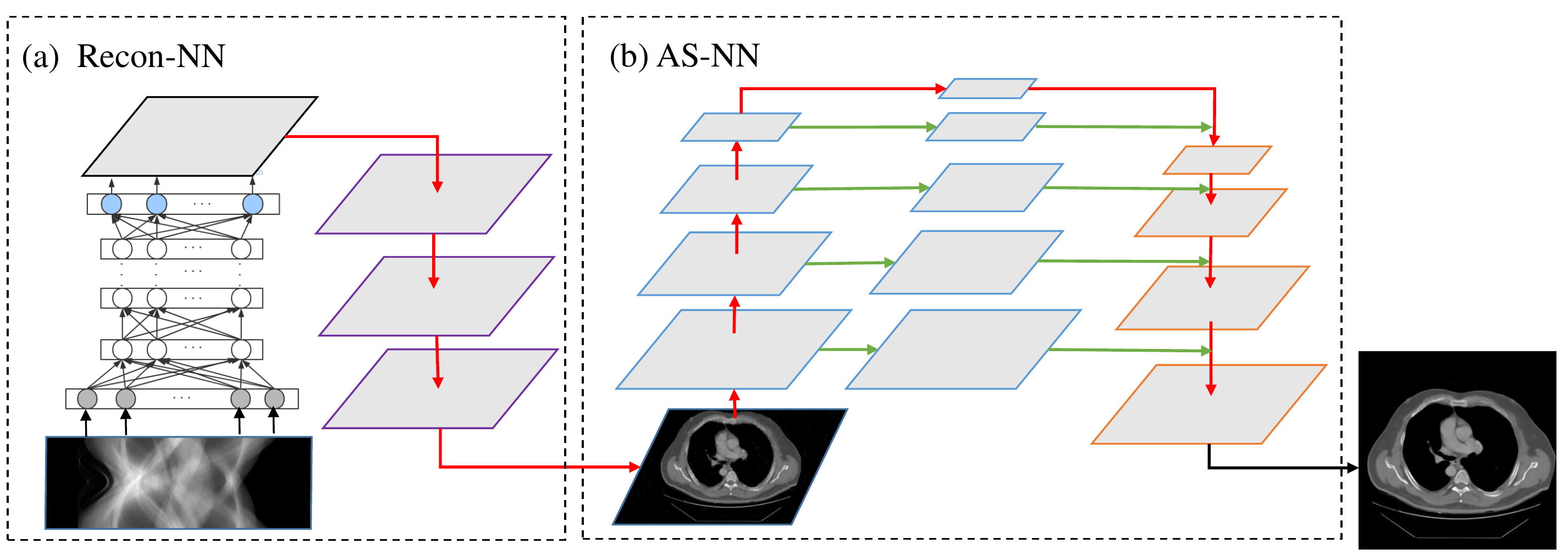}
%\put (-495, 86){\large (a) Recon-NN}
%\put (-218, 86){\large (b) AS-NN}
\caption{The whole architecture and dataflow chart of the FAR-net, which include two parts of CT imaging processing. (a) Recon-NN for image reconstruction strategy, (b) AS-NN for artifact suppressing and de-noising.}\label{All_net}
\end{figure*}

In summary, as shown in Fig. \ref{All_net}, the FAR-net is an end-to-end network, which is composed of Recon-NN and AS-NN. Though pre-training is not required for the FAR-net,  it could be  considered as two steps: in the first step (Recon-NN), the FAR-net predict the CT image from input sparse-views projection data directly with artifacts and noise. Then the CT image are processed by the second step of FAR-net (AS-NN) to improve the image quality.

\section{Numerical experiments}\label{NE}
In this section, various experiments are carried to evaluate the FAR-net as well as the matrix factorization.
\subsection{Validation of matrix factorization}\label{A_1}
We first perform some numerical experiments to validate the Equation (\ref{AWWW}). In order to simplify problem, we remove the convolution layers of Recon-NN in the testing (only FC network remained). However, it is difficult to estimate directly whether the predicted matrix  which is defined as $\widetilde{A}$  is inverse of $A$. So we calculate $E \approx W_k\cdots W_2\cdot W_1\cdot A=\widetilde{A}A$ as the evaluation objective, which should be equal or approximately equal to identity matrix ($I$). 
To train and validation the FC network (the simple version of Recon-NN), we have to prepare a large dataset comprising pairs of input data and label data. Firstly, we generate a nonnegative sparse matrix$A_{N\times N}$ randomly, of which the sparsity is $\frac{2\sqrt N}{N}$. Then, we select 100 images from Pascal VOC \cite{VOCdata} and resize them to $N\times N$. Each row ($X_i$) of image can be regarded as ground truth data and $b_i (=A\cdot X_i)$ can be regraded as input data. Hence, the total number of training and testing data pairs is $100\times N$. It is should be noticed that the operator is  only a matrix multiplication rather than radon transform in this experiment. The configuration of the FC network is displayed in Table \ref{FC_network_prama}.
\begin{table}
  \centering
  \caption{The configuration of FC network}\label{FC_network_prama}
  \setlength{\tabcolsep}{1.5mm}{
  \begin{tabular}{c| c c}
  %\toprule
  \hline \hline
  \multirow{3}*{hardware}&CPU&E5-2620 v3 $\times$ 2\\
  %\cline{2-3}
  & GPU&\tabincell{c}{NVIDIA GTX1080Ti\\} \\
  %\cline{2-3}
  &memory&512GB \\
  \hline
  \multirow{3}*{software}&language&python3.6\\
  %\cline{2-3}
  &deep learning frame&pytorch0.4.0\\
  %\cline{2-3}
  &CUDA version& cuda9.1, cudnn7.1\\
  \hline
  \multirow{3}*{hyper parameters}&batch size&3000\\
  %\cline{2-3}
  &learning rate &0.0001\\
  %\cline{2-3}
  &optimizer&adam\\
  \hline
  \end{tabular}}
\end{table}

\begin{figure}
  \centering
  \includegraphics[width=2.8in]{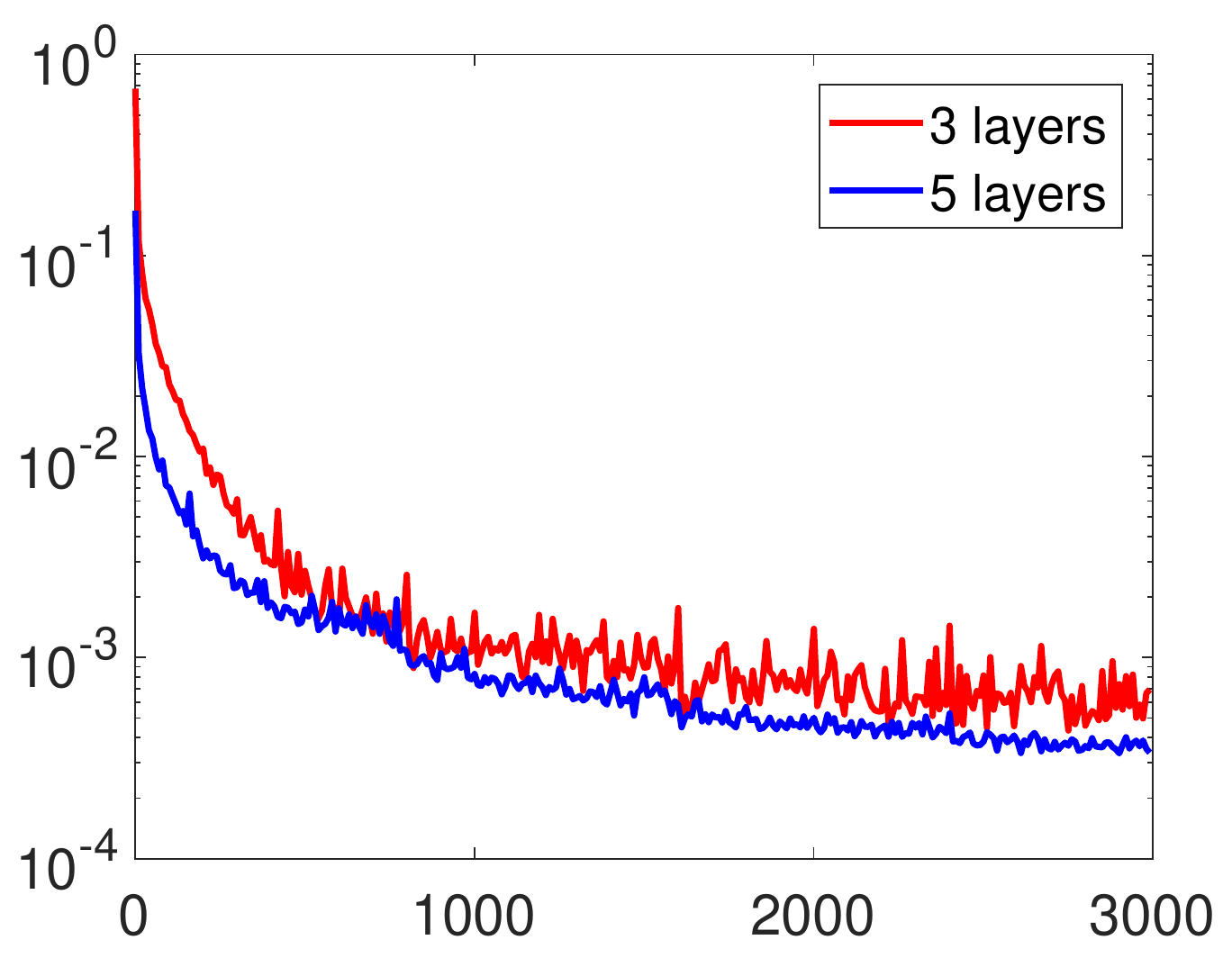}
  \put (-110, -7){\large epoch}
  \put (-210, 60){\begin{turn}{90} \large loss value \end{turn}}
  \caption{MSE Loss function update with different depth of FC network. The convergence and loss value of five layers of network are significantly lower than those of three layers of network.}
  \label{depth_loss}
\end{figure}

\begin{figure}[htbp]%
\centering
\includegraphics[width = 2.55in]{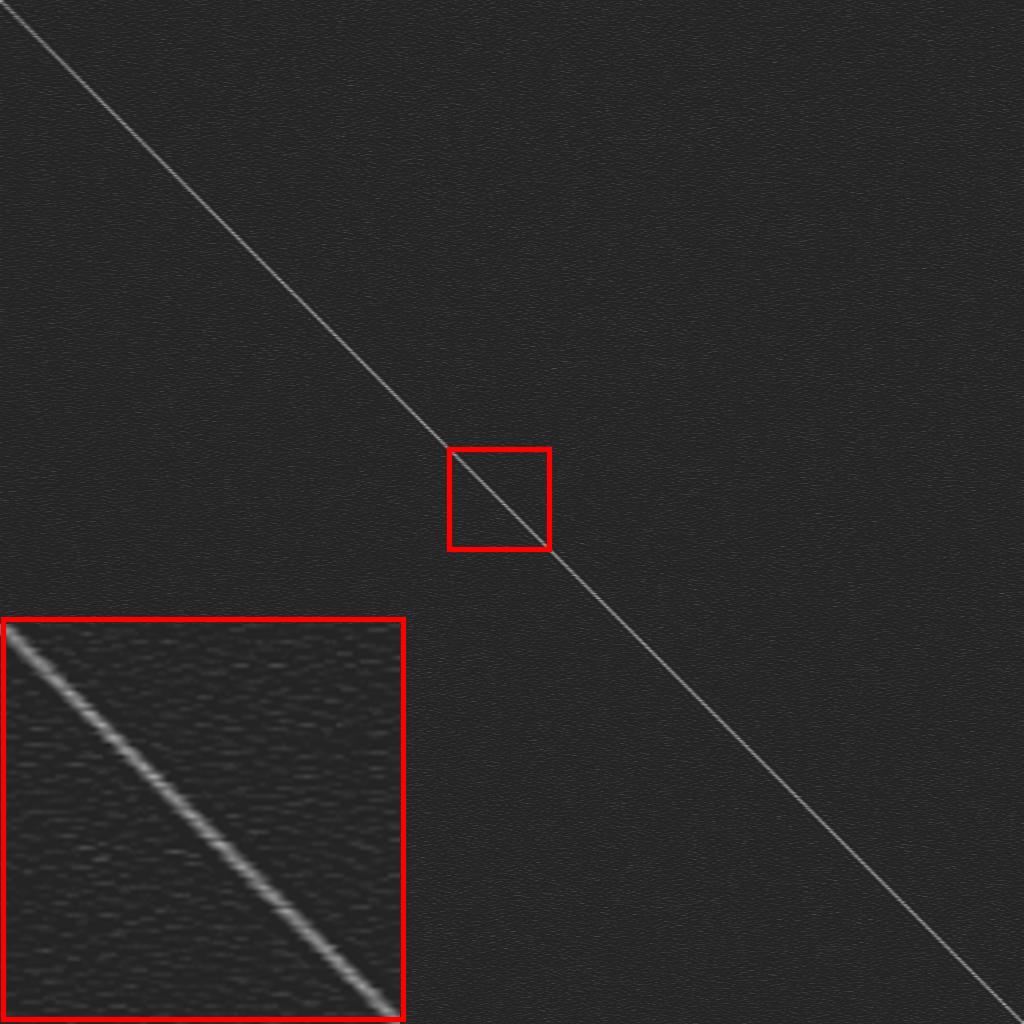}
%}
\caption{A representative result of matrix multiplied by $A$ and the inverse matrix trained by FC network. The dimension of  $A$ is $1024\times 1024$ and display windows is [-0.1, 0.6].}
\label{unit matrix}
\end{figure}

\begin{figure}
  \centering
  \includegraphics[width=3in]{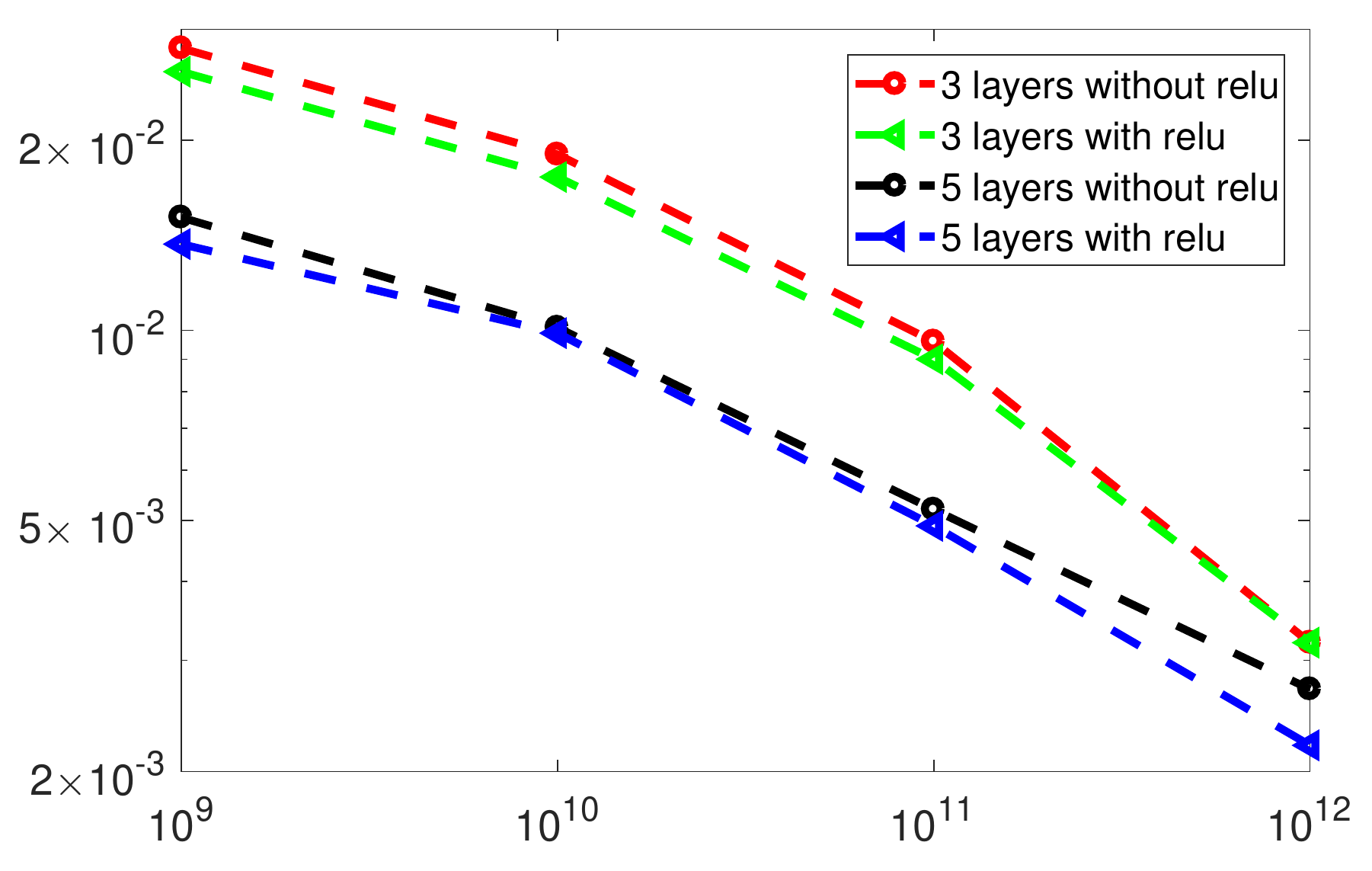}\\
  \caption{The curve of NMAD between $E$ and $I$ with the dimension of the matrix $A$. The X axis represents the dimension of matrix and different curves represent different situations of networks}
  \label{psnr_dimA}
\end{figure}

In this testing, there are 4 aspects are studied for the impact of FC network in FAR-net: the depth of network, whether using nonnegative constraint (rectified linear unit activation function), the times of back-propagation and the dimension of $A$. Hence, we testing different matrix dimension and configurations of FC network. The results show that the inverse of the SNM is able to trained by the FC neural network with several smaller layers, and the dimension of corresponding  matrices are much smaller than the inverse matrix. Furthermore, we can conclude some rules as follows:
\begin{itemize}
  \item Fig. \ref{depth_loss} shows loss curves with different number of middle layers. It is obviously that increasing the number of middle layers can not only accelerate convergence and improve accuracy, but also keep smoothing of loss curve.
  
  \item The results of weight matrix multiplied by matrix $A$ ($E = W_k\cdots W、_2\cdot、 W_1\cdot A$) with different parameters are shown in Fig. \ref{unit matrix}, where the dimension of $A$ is $N=1024$. It is illustrated that the matrix $E$ are approximately equal to the identity matrix $I$ (most of the nonnegative value are  distributed on the diagonal).
  
  \item From the Table \ref{psnr_E}, under the NMAD, it is noticed that the factorization can achieve better performance with nonnegative constraint  which is a very similar conclusion with the SNMF.
  
  \item Fig. \ref{psnr_dimA} shows that the variation of NMAD with the different dimension of the matrix $A$. We can see that NMAD decreases as the matrix dimension increases. In fact, when the reconstructed image size is $512\times 512$, the dimension of matrix $A$ can be regarded as $2^{9} \times 2^{9} = 2^{18}$ for reconstruction problem.

\end{itemize}

\begin{table}[htbp]
  \centering
  \caption{NMAD between matrix $E$ and identity matrix}\label{psnr_E}
  \setlength{\tabcolsep}{1.5mm}{
    \begin{tabular}{c c c c c c c}
    \toprule
    \hline
     \makecell{matrix\\ dimension}& epoch & \makecell{3 layers \\ without relu} & \makecell{5 layers \\ without relu} & \makecell{3 layers \\with relu} &\makecell{5 layers \\with relu}  \\
    \hline
    \multirow{3}*{$512\times 512$}& 1000  & 0.0331   & 0.0257& 0.0182& {\bfseries 0.0143} \\
    & 2000  & 0.0294   & 0.0254& 0.0159   & {\bfseries 0.0142}\\
    & 3000  & 0.0280   & 0.0257& 0.0151& {\bfseries 0.0137}\\
    \hline
    \multirow{3}*{$1024\times 1024$}& 1000  & 0.0237   & 0.0210& 0.0124   & {\bfseries 0.0114}\\
    & 2000  & 0.0203   & 0.0198& 0.0109& {\bfseries 0.0108}\\
    & 3000  & 0.0190   & 0.0175& 0.0101  & {\bfseries 0.0099}\\
    \hline
    \multirow{3}*{$2048\times 2048$}& 1000  & 0.0116  & 0.0127 & 0.0067   & {\bfseries 0.0062}\\
    & 2000  & 0.0093   & 0.0113& 0.0061  & {\bfseries 0.0052}\\
    & 3000  & 0.0096   & 0.0090& 0.0052& {\bfseries 0.0049}\\
    \hline
    \multirow{3}*{$4096\times 4096$}& 1000  & 0.0035  & 0.0034 & {\bfseries 0.0024} & {\bfseries 0.0024}\\
    & 2000  & 0.0032   & 0.0032& 0.0028  & {\bfseries 0.0024}\\
    & 3000  & 0.0032   & 0.0032& 0.0027   & {\bfseries 0.0022}\\
    \hline
    \end{tabular}}
\end{table}

\subsection{Sparse-views CT reconstruction}\label{CT_exper}
In this subsection, the FAR-net is evaluated with the medical image dataset and compared with conventional approaches for sparse-views CT. 
\subsubsection{Dataset and configuration}
 In the numerical experiments for sparse-views CT, we select the TCGA-ESCA cancer CT image dataset \cite{dataset} as the test object. We chose 4302 images from the dataset with the size of $512\times 512$ (pixels), which are regarded as the ground truth of training dataset. And the input dataset of neural network are generated by using the ground truth images to simulate parallel beam projection. The parameters of parallel beam CT projection are set as follows: the total number of views is 60, of which the interval is 3 degrees and there are 600 rays for each view. Otherwise, the testing dataset are generated in the same way and are not included in the training dataset. The  configuration of whole processing of the FAR-net are displayed in Table \ref{whole_network_prama}.
\begin{table}
  \centering
  \caption{The experiments configuration of whole processing of our proposed network}\label{whole_network_prama}
  \setlength{\tabcolsep}{1.5mm}{
  \begin{tabular}{c| c c}
  %\toprule
  \hline \hline
  \multirow{3}*{hardware}   &CPU      &$2\times$ E5-2620 v3\\
                            &GPU  &9$\times$ NVIDIA GTX1080Ti\\
                            &memory          &512GB \\
  \hline
  \multirow{3}*{software}   &operation system &windows 10\\
  &deep learning framework&pytorch0.4.0\\
  &GPU plantform& cuda9.1, cudnn7.1\\
  \hline
  \multirow{4}*{hyper-parameter}&batch size &1\\
  &learning rate &$10^{-4}\sim 10^{-6}$\\
  &optimizer&adam\\
  &training iterations & 600 epochs\\
  &training time & 68h35m\\
  \hline
  \end{tabular}}
\end{table}

\subsubsection{results}
\begin{figure*}[htbp]%
%\centering
\includegraphics[width = 7.2in]{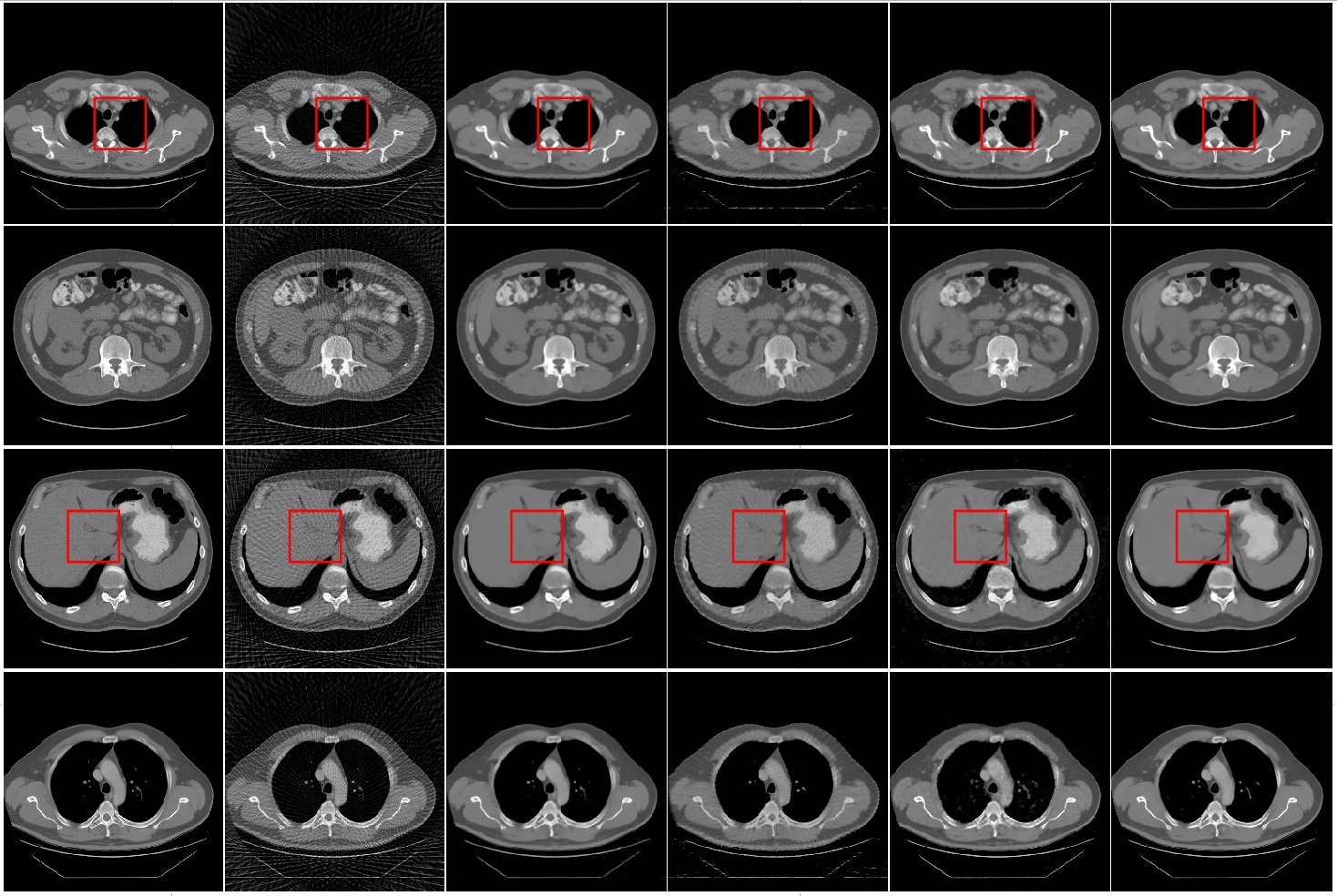}
\put (-482, -9){ (a)}
\put (-397, -9){ (b)}
\put (-310, -9){ (c)}
\put (-220, -9){ (d)}
\put (-135, -9){ (e)}
\put (-50, -9){ (f)}
\caption{Parts of reconstruction results from the test dataset without noise: (a) ground truth, (b) FBP algorithm, (c) TV based, (d) FBP+U-net, (e) Recon-NN, (f) FAR-net. The display windows are [700Hu, 1500Hu].}
\label{image_res}
\end{figure*}

\begin{figure*}[htbp]%
\centering
\includegraphics[width = 7.2in]{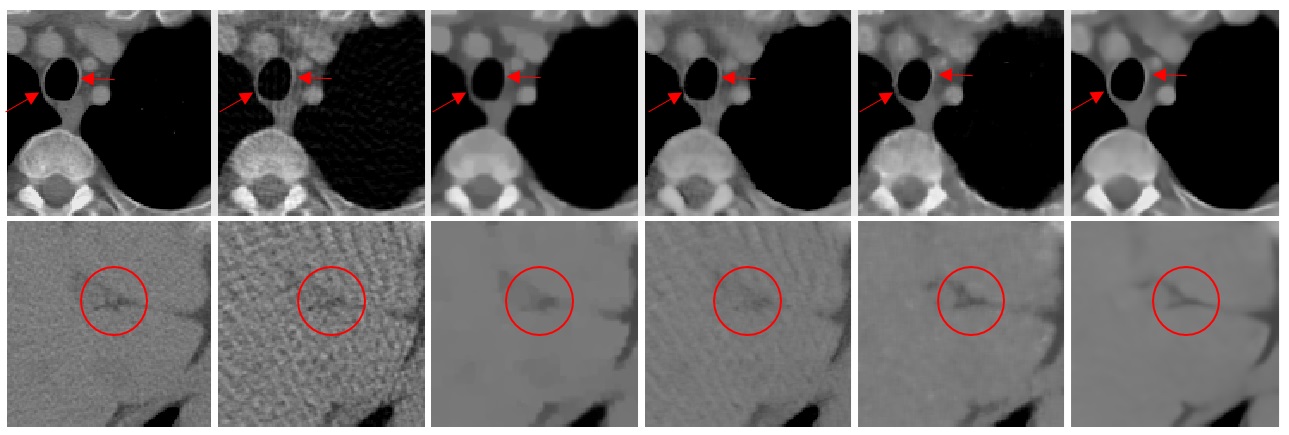}
\put (-482, -9){ (a)}
\put (-397, -9){ (b)}
\put (-310, -9){ (c)}
\put (-220, -9){ (d)}
\put (-135, -9){ (e)}
\put (-50, -9){ (f)}
\caption{Zoomed in region of interest (ROI) marked by the red box in Fig.\ref{image_res}. (a) ground truth, (b) FBP algorithm, (c) TV based, (d) FBP+U-net, (e) Recon-NN, (f) FAR-net. The arrows indicate two locations with significant visual differences.}
\label{ROI_noise_free}
\end{figure*}

\begin{figure}[htbp]%
\centering
\includegraphics[height = 2.2in, width = 3.2in]{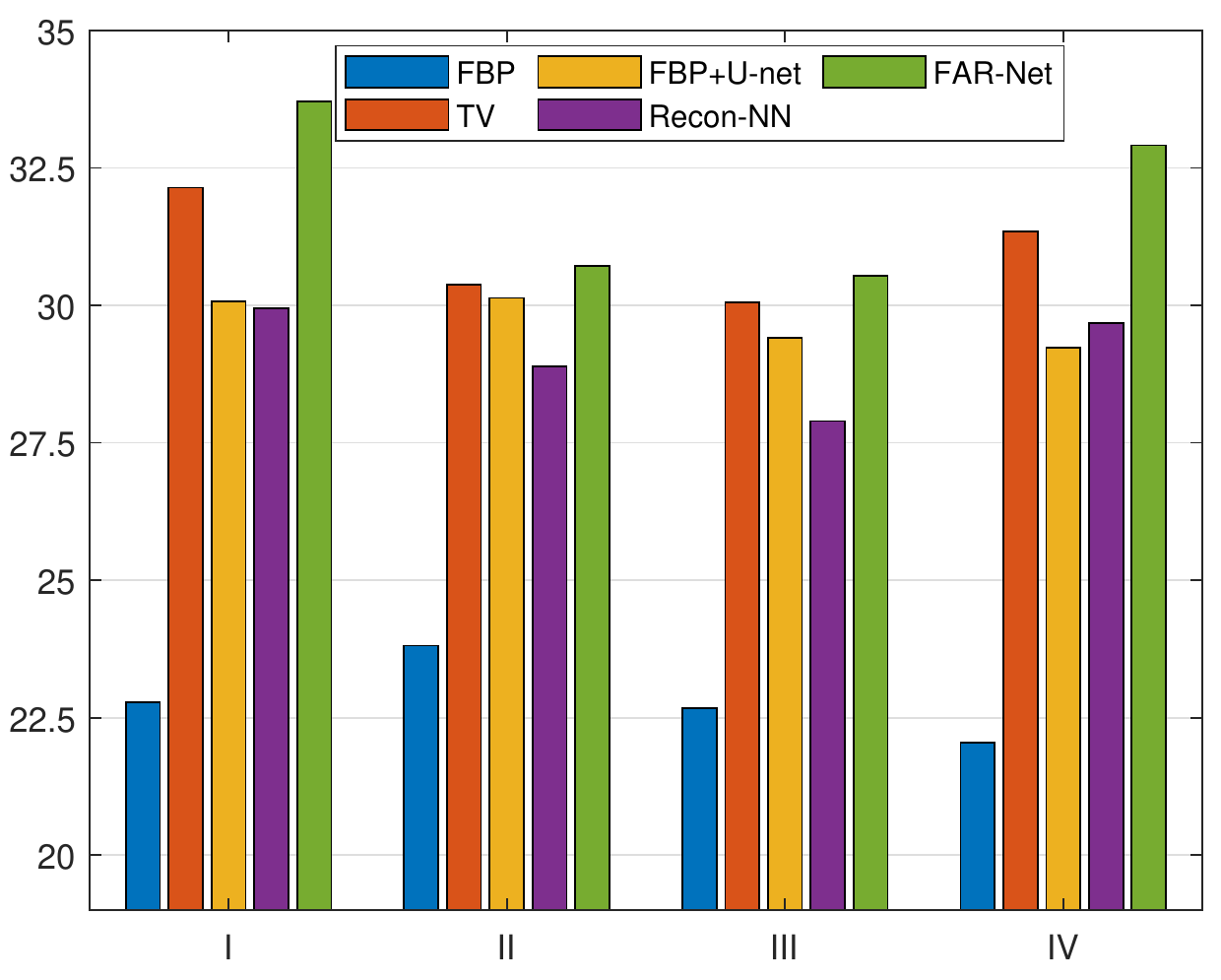}
%\put (-180, 160){PSNR for Images shown in Fig.\ref{image_res} }
\put (-240, 70){\begin{turn}{90} PSNR \end{turn}}
\caption{PSNR performance comparison of different algorithms for the images shown in Fig.\ref{image_res}. }
\label{PSNR_noise_free}
\end{figure}

\begin{figure}[htbp]%
\centering
\includegraphics[height = 2.2in, width = 3.2in]{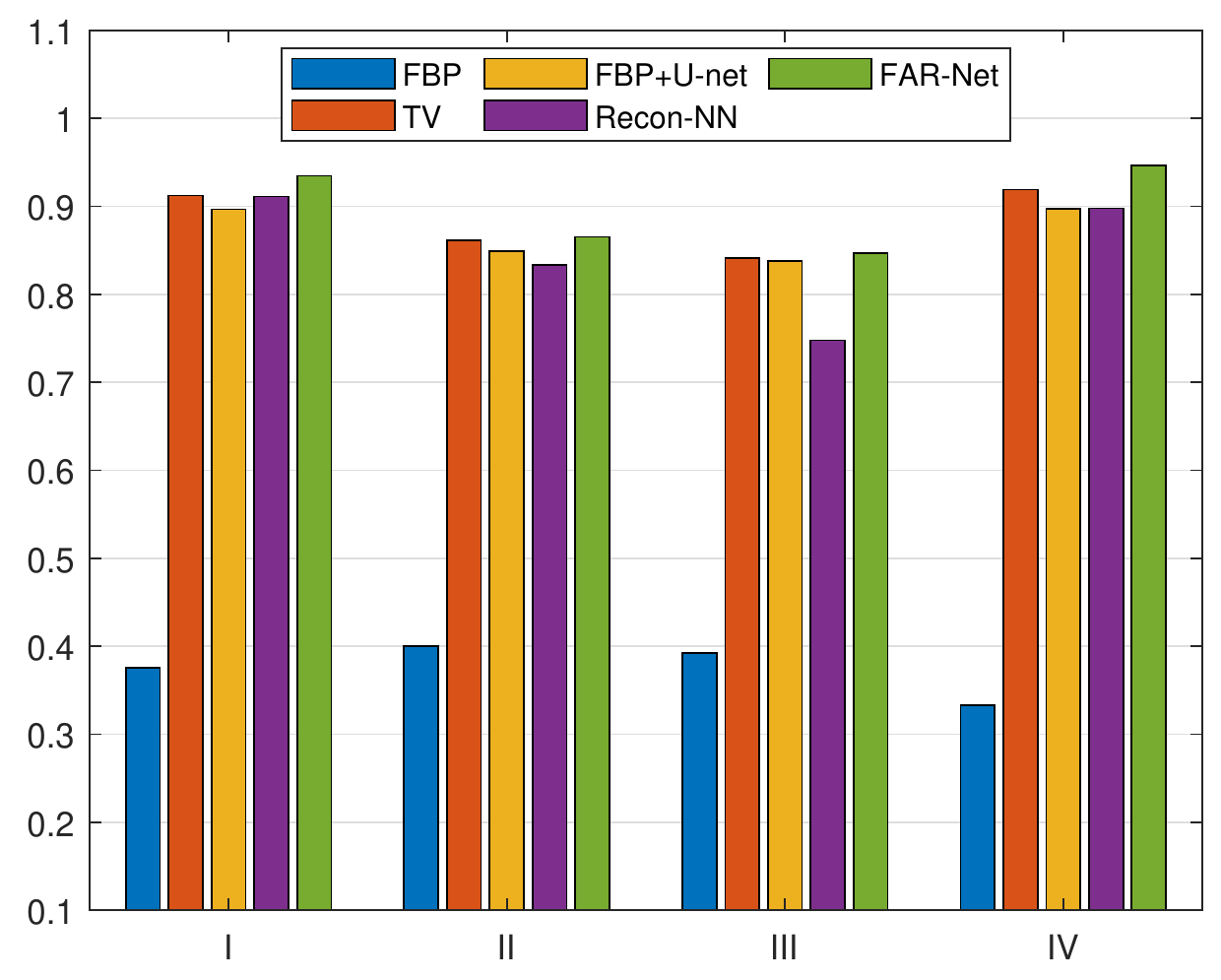}
\put (-240, 70){\begin{turn}{90} SSIM \end{turn}}
\caption{SSIM performance comparison of different algorithms for the images shown in Fig.\ref{image_res}. }
\label{SSIM_noise_free}
\end{figure}

\begin{figure*}[htbp]%
%\centering
\includegraphics[width = 7.2in]{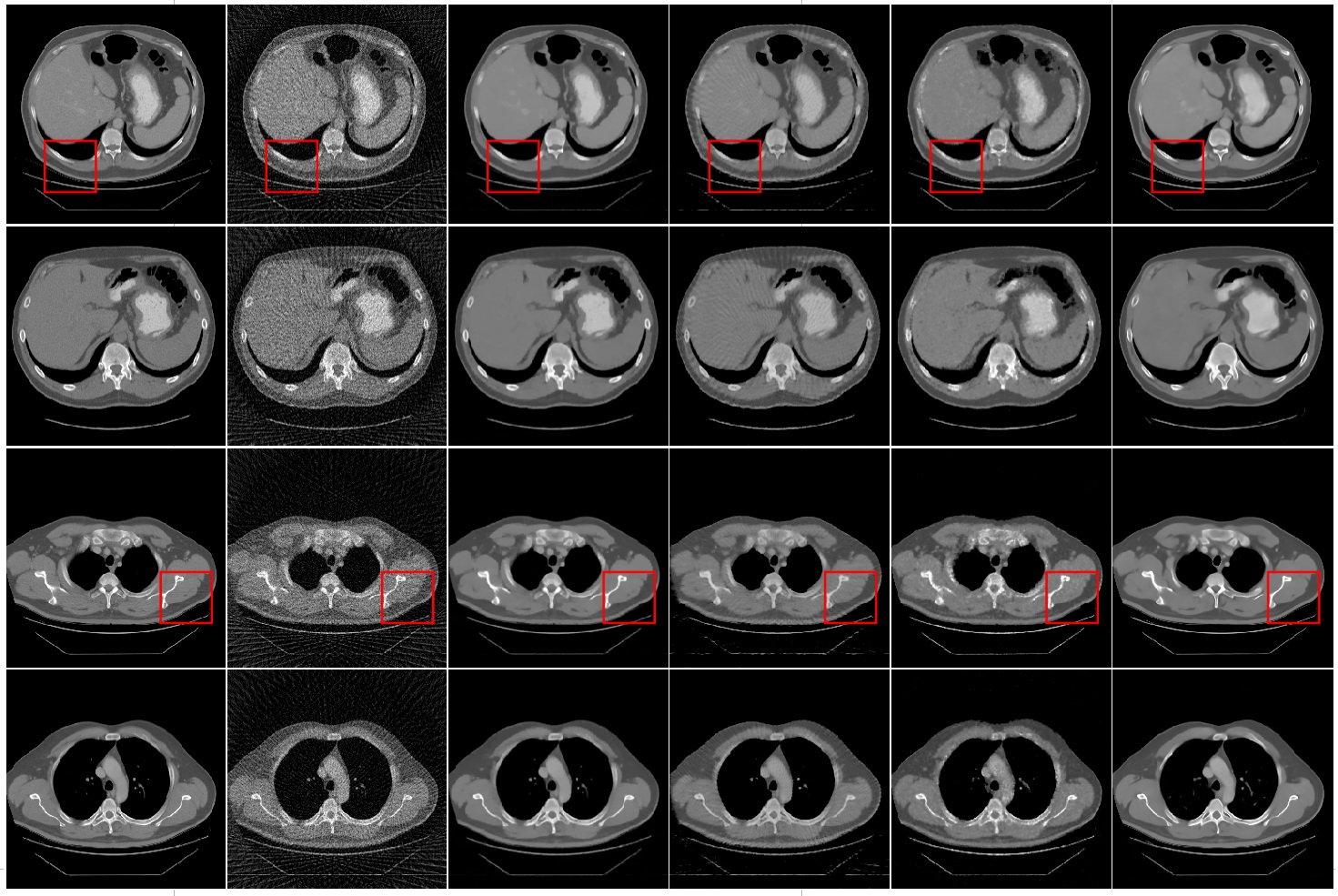}
\put (-482, -9){ (a)}
\put (-397, -9){ (b)}
\put (-310, -9){ (c)}
\put (-220, -9){ (d)}
\put (-135, -9){ (e)}
\put (-50, -9){ (f)}
\caption{Reconstruction results from the test dataset with noisy: (a) ground truth, (b) FBP algorithm, (c) TV based, (d) FBP+U-net, (e) Recon-NN, (f) FAR-net. The display windows are [700Hu, 1500Hu].}
\label{image_res_noise}
\end{figure*}

\begin{figure*}[htbp]%
\centering
\includegraphics[width = 7.2in]{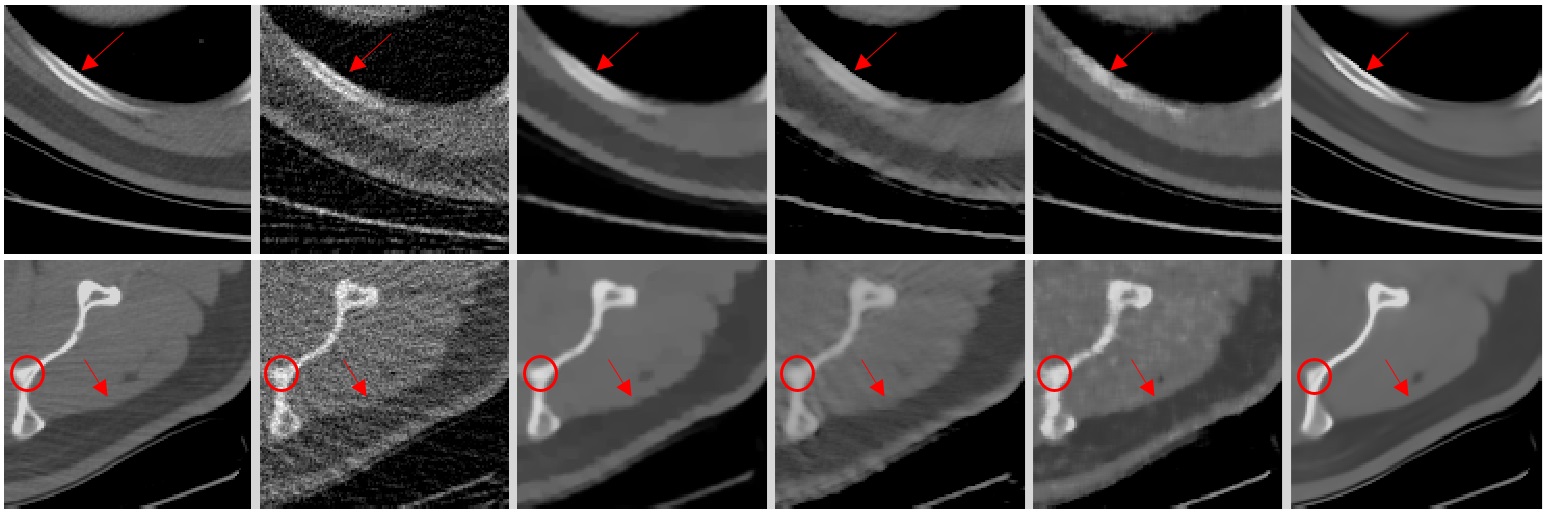}
\put (-482, -9){ (a)}
\put (-397, -9){ (b)}
\put (-310, -9){ (c)}
\put (-220, -9){ (d)}
\put (-135, -9){ (e)}
\put (-50, -9){ (f)}
\caption{Zoomed in region of interest (ROI) marked by the red box in Fig.\ref{image_res_noise}. (a) ground truth, (b) FBP algorithm, (c) TV based, (d) FBP+U-net, (e) Recon-NN, (f) FAR-net. The arrows indicate two locations with significant visual differences.}
\label{ROI_noisy}
\end{figure*}

\begin{figure}[htbp]%
\centering
\includegraphics[height = 2.2in, width = 3.2in]{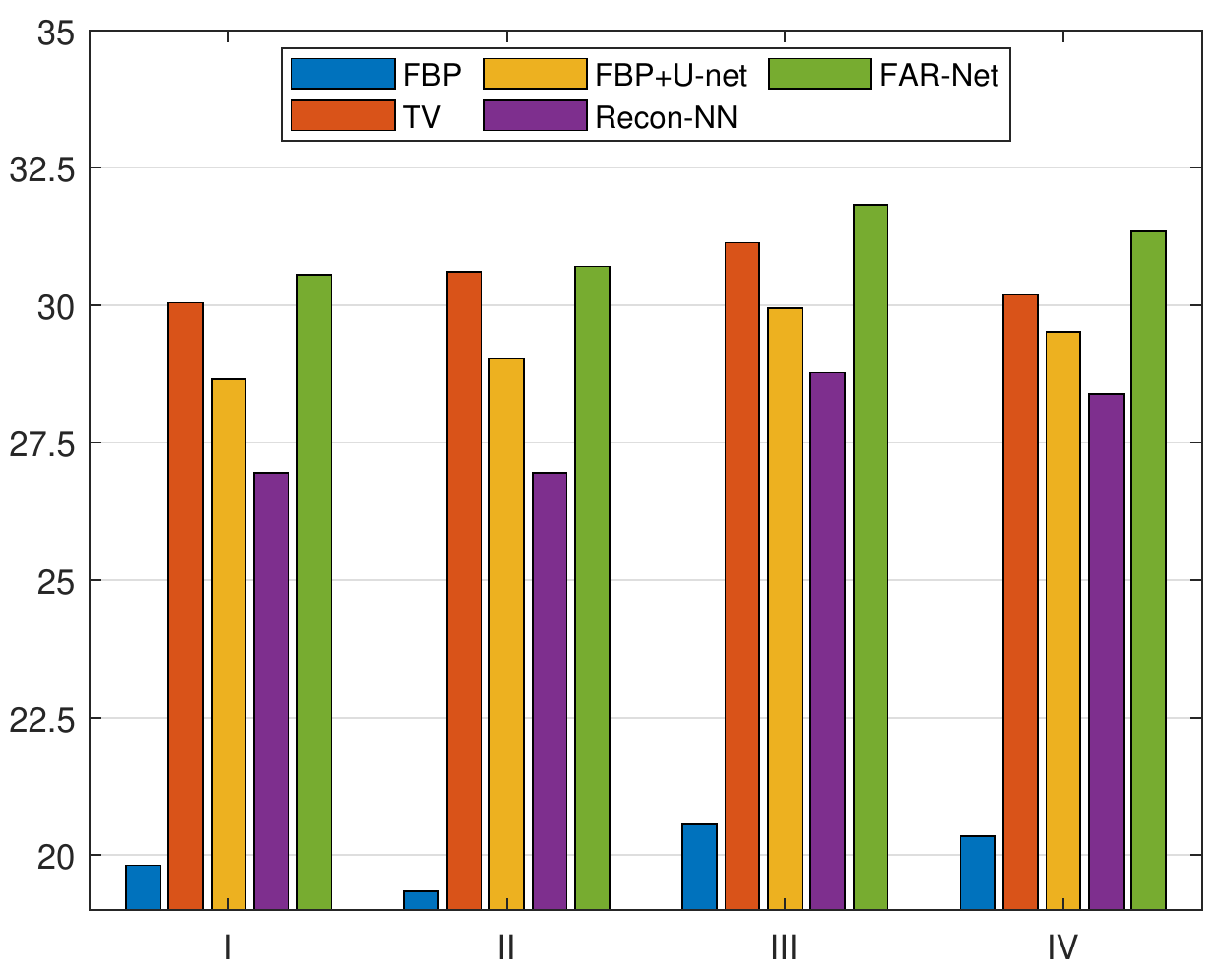}
\put (-240, 70){\begin{turn}{90} PSNR \end{turn}}
\caption{PSNR performance comparison of different algorithms for the images shown in Fig.\ref{image_res_noise}. }
\label{PSNR_noisy}
\end{figure}
\begin{figure}[htbp]%
\centering
\includegraphics[height = 2.2in, width = 3.2in]{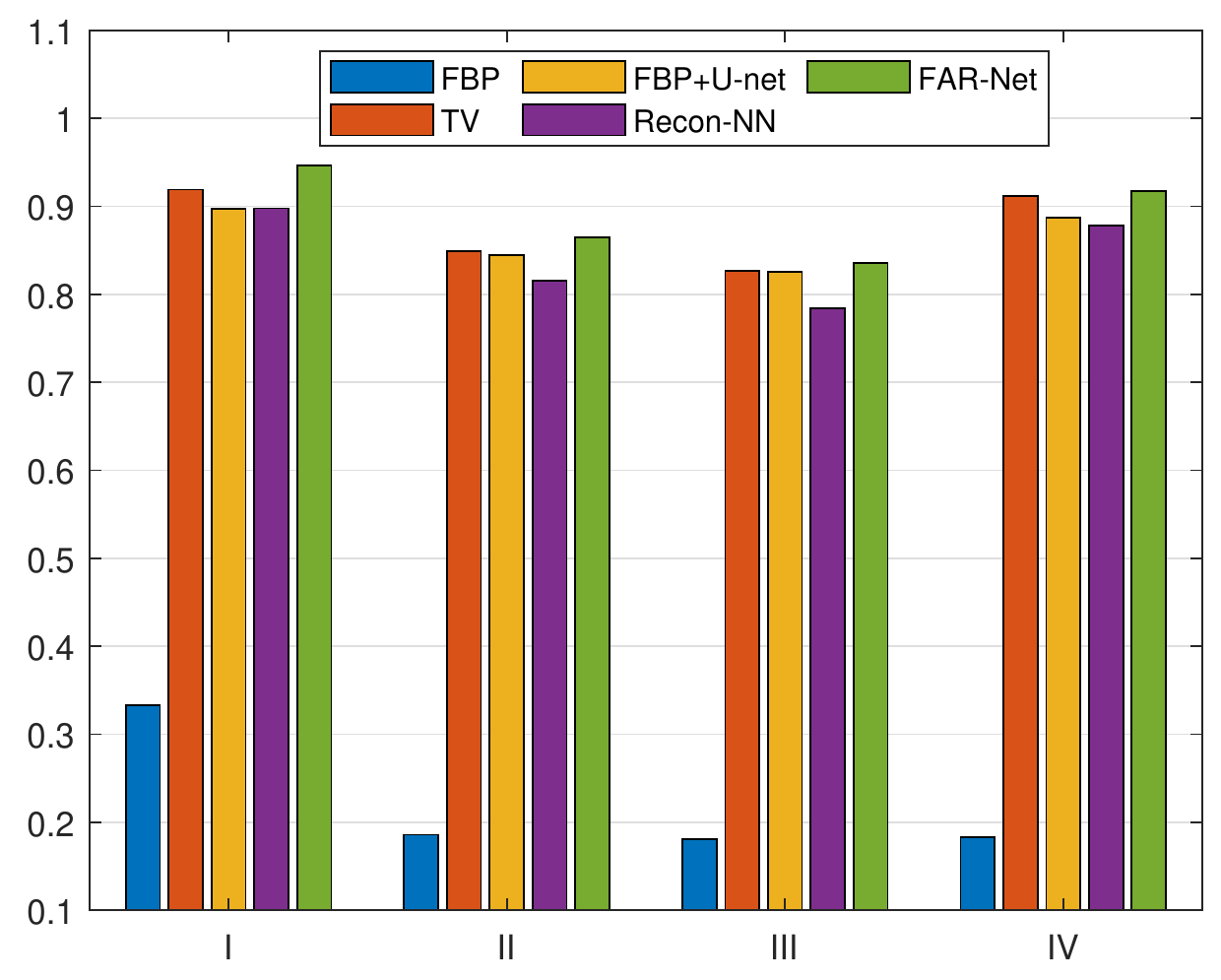}
\put (-240, 70){\begin{turn}{90} SSIM \end{turn}}
\caption{SSIM performance comparison of different algorithms for the images shown in Fig.\ref{image_res_noise}. }
\label{SSIM_noisy}
\end{figure}

The images in testing dataset are predicted by the FAR-net with sparse-views projection data. For comparison, FBP algorithm, optimization-based algorithm and deep learning methods are also utilized to  reconstruct images from the sparse-views projection data. In the optimization-based method, the regularization term of  the objective function is the Anisotropic TV (Rudin-Osher-Fatemi model) of image and solving algorithm is Split Bregman method, 
\begin{equation}
u = argmin (|d_xu|+|d_yu|+\frac{\mu}{2}||Au-f||_2^2)
\end{equation}
The regularization parameters $\mu$ for ROF is set to 100 and  $\lambda$ for the split-Bregman is set to 50. Moreover, U-net is chosen as deep learning based method for comparing the proposed method. The  images reconstructed with sparse-view projection by FBP  and the corresponding reference are set as input and label to training U-net, respectively. Hyper-parameter such as learning rate, batch size are consistent with AS-NN. In the rest of the paper, we use FBP+U-net to denote this approach.

Fig. \ref{image_res} lists some example images which are reconstructed from 60 views with FBP, TV based method, FBP+U-net, Recon-NN and the FAR-net, respectively. It is obviously that images reconstructed from FBP suffer with heavy artifacts caused by under-sampling projection. The quality of reconstructed image is degraded by streak artifacts distributed across the whole image. All these methods suppressed image artifacts and improve image quality to various levels.The TV based method and FAR-net take more efficiently for suppressing the streak artifacts and achieving better performance in image quality. However, as shown in the Fig. \ref{image_res} (c), TV based method suffered from a blocky effect and also smoothened some important small structures, such as image edge.

The example results of FBP+U-net method are shown in Fig. \ref{image_res} (d). It is demonstrated that FBP+U-net removes major artifacts but some smaller streak artifacts are still available.The results of FAR-net are shown in the last column of Fig. \ref{image_res}. It is illustrated that the reconstruction image has no obvious blocky effect and has a best performance than other methods in low contrast region.

Fig.\ref{ROI_noise_free} shows the zoomed in region of interest (ROI) marked by in Fig. \ref{image_res}. As indicated by the red arrows and circles, it is noticed that the FBP, TV based and FBP+U-net make the edges and low contrast region blurred or distorted, where FAR-net method can still maintain the structure. 

%This is not surprising since the TV based algorithm, just one parameter, balances the fidelity and the regularization. Although the neural network has strong mapping ablity, the FBP+U-net method feeds into the neural network is low quality images, which results in artifacts remaining in the reconstructed images and low contrast such as bone structure. Ours method  consists of two strategies, the first strategy (Recon-NN) have obtained higher quality reconstruction results. Hence the second strategy can further obtain high quality images.%
The proposed method is not only be superior to conventional method (TV based) but also better than deep learning based post-processing method (FBP+U-net). Comparing with FBP+U-net method, ours method  consists of two strategies and the first strategy (Recon-NN) have obtained higher quality reconstruction results than FBP. Hence the second strategy can further obtain high quality images. The image quality indexs such as PSNR and SSIM are displayed in Fig. \ref{PSNR_noise_free} and Fig \ref{SSIM_noise_free}, respectively.  The proposed FAR-net performs the best. It is also proved that AS-NN achieves better experimental results based on better initial value mapping.

Furthermore, we test these methods in noise situations. Poisson noise specified by the incident intensity, denoted as $I_0$, would alwayss be added to the raw data, i.e.
\begin{equation}
p_{noisy} = -ln (\frac{Poissrnd (I_0\times e^{-p})}{I_0})
\end{equation}
where $p$ and $p_{noisy}$ denote noise-free and noisy projection data, respectively. Here, we simulated Poisson noise with incident intensity $I_0 = 1.0\times 10^6$ . Fig. \ref{image_res_noise} show the results of different methods. In general, Both FBP+U-net and the FAR net are able to  suppress the noise and remove the steak artifacts at the same time. However, the blocky effect is still available in the image obtained by FBP+U-net.

Fig. \ref{ROI_noisy} are the regions of interest marked by red arrows and circles in Fig. \ref{image_res_noise}. As indicated by the red arrows and cicrle, it is clearly shown that images predicated by FAR net have obvious advantages both in contrast and edge preservation.   The blurred structures and over-smoothness at the edge are appeared in the images reconstructed by TV based method.  For the FBP+U-net method, similar to noise free conditions, some artifacts are still remained and low contrast details are submersed. Furthermore, The above results are validated by image quality index. As shown in Fig. \ref{PSNR_noisy} and Fig. \ref{SSIM_noisy}, the PSNR and SSIM are calculated.  According to the results above, the proposed FAR-net can predict images with better performance both in vision and quality. Furthermore, the deep learning method has a higher efficient for image reconstruction. For example, the FAR-net only consume 0.0845 seconds to predict a image from projection data which is much less than conventional method such as TV based. 

% which can see the boundary marked by red arrow. For the FBP+U-net method,  the reconstructed image still has streak artifacts caused by sparse samplings, and the contrast of reconstructed image is not satisfied, e.g., the CT value in the part of bone is lower than ground truth. For the proposed method, the FAR-net has a better performance on the edges, e.g., the sharp edges are maintained well.
%

%Table \ref{PSNR_figureRes} compares the SSIM and PSNR of sparse-views reconstruction using FBP, TV based, FBP+U-net, Recon-NN and FAR-net. The FAR-net generally performs best in terms of SSIM and PSNR. The reason for this is that the powerful function of convolution layer which enhancement the two distance measures for reconstructed images. With a detail image, the network can learn where the prominent streaks are in the uncorrected image. Moreover, for the deep learning method, although the training stage is time consuming, the time cost of predication is much fewer than the TV based and other iterative reconstruction methods. The time cost of reconstruction shown in Table \ref{use_time}. Once the parameters of network are trained, the FAR-net only spend 0.0845s from radon domain to image domain and reconstruction is efficient than conventional method such as TV based. 

\section{Conclusion}\label{summary}
In this paper, we propose a neural network to map the CT reconstruction processing which is able to predict CT images from sparse-views projection data automatically. Different from most of the relevant works, which treated the neural network as black boxes, the FAR-net was directly motivated by sparseness non-negative matrix factorization and all the parameters are learned from training samples rather than pre-calculated. Furthermore, the whole net is divided into two sub structures including Recon-NN and AS-NN, which are a two-stage training strategy on single object images and multi-object images. This strategy makes the network deeper and realize a coarse-to-fine learning process and the size of whole net are optimized from $O (n^{4})$ to $O (n^{3})$ which could be distributed in a single workstation with multi-GPU. Numerical experiments show that the FAR-net can be effectively introduced into the reconstruction process and has shown the outstanding advantages in terms of noise suppression, artifact reduction, edge and feature preserving. Comparing to  the conventional methods,  the FAR-net has demonstrated a superior performance over in both image quality and computational efficiency.

\bibliographystyle{IEEEtran}
\bibliography{IEEEabrv,mybib}

%\begin{IEEEbiography}{Genwei Ma}
%Biography text here.
%\end{IEEEbiography}
%
%% if you will not have a photo at all:
%\begin{IEEEbiographynophoto}{Yining Zhu}
%Biography text here.
%\end{IEEEbiographynophoto}
%
%% insert where needed to balance the two columns on the last page with
%% biographies
%%\newpage
%
%\begin{IEEEbiographynophoto}{Xing Zhao}
%Biography text here.
%\end{IEEEbiographynophoto}
\end{document}